# VIPE: A NEW INTERACTIVE CLASSIFICATION FRAMEWORK FOR LARGE SETS OF SHORT TEXTS - APPLICATION TO OPINION MINING

W. Siblini[1,2*], F. Meyer[1] and P. Kuntz[2]
[1]Orange Labs Lannion
2 av. Pierre Marzin - 22 300 Lannion, France
<firstname>.<lastname>@orange.com

[2]Laboratoire des Sciences et du Numerique de Nantes
Site Polytech Nantes – 44 300 Nantes cedex, France
<firstname>.<lastname>@univ-nantes.fr

**ABSTRACT**

This paper presents a new interactive opinion mining tool that helps users to classify large sets of short texts originated from Web opinion polls, technical forums or Twitter. From a manual multi-label pre-classification of a very limited text subset, a learning algorithm predicts the labels of the remaining texts of the corpus and the texts most likely associated to a selected label. Using a fast matrix factorization, the algorithm is able to handle large corpora and is well-adapted to interactivity by integrating the corrections proposed by the users on the fly. Experimental results on classical datasets of various sizes and feedbacks of users from marketing services of the telecommunication company Orange confirm the quality of the obtained results.

**Keywords**: opinion mining, interactive learning, multi-label classification

## 1 INTRODUCTION

Opinion mining aims at extracting opinions expressed in user-generated contents (Gundecha [1]) and at assisting managers to discover the « true Voice of the Customer » (Power [2]). It helps to determine the level of excitement generated by technical and commercial offers, to identify the different client perspectives, and to gather fresh information from social media. Interest from both industry and academia have stimulated numerous works during the last decade as confirmed by a very recent bibliometric analysis (Piryani [3]).

Roughly speaking, opinion mining, often associated to sentiment analysis, combines various computational techniques from natural language processing, data mining and machine learning. Web surveys, forums and social networks are the main sources for opinion analysis in companies and consequently the major part of available data is heterogeneous, unstructured and very noisy. Hence, the acquisition of useful information for decision-making is hard and raises important research challenges. Significant efforts have been devoted to the development of automatic dedicated algorithms (Ravi [4]). But, despite important progress, in particular in textual data processing, some researchers and practitioners do not hesitate to argue that for fine-grained opinion mining manual annotation remains unavoidable (Sun [8]) and that user involvement is crucial (Keim [5]). The development of new machine learning methods capable of working collaboratively with humans to jointly analyse complex data sets is cited among the major emerging trends in a recent review on machine learning published in Science (Jordan [6]). As defined by Amershi [7] « interactive machine learning is a process that involves a tight interaction loop between a human and a machine learner, where the learner iteratively takes input from the human, promptly incorporates that input, and then provides the human with output impacted by the results of the iteration ».

---

[*] Corresponding Author

In this paper, we present a new interactive learning framework, called VIPE (Visual Interactive and Personalized Exploration of Data), for the classification of large sets of short texts originated from results of opinion polls on the Web, call center verbatim, technical forums and Twitter. Users initially classify a limited set of texts manually with a set of pre-defined labels in which they are interested. For instance, let us consider a telecommunication company –at the origin of the use case presented in this paper- where agents analyse opinions concerning its products. An agent can annotate the tweet « this is the best 4G network! » with the labels « positive opinion », « efficiency » and « innovation ». Then, from this restricted tagged text set, a learning algorithm predicts the labels of the remaining texts of the corpus and the texts most likely associated to a selected label. In the process, the short texts are initially represented by traditional n-grams of words without any additional linguistic information and the learning stage is performed by a fast matrix factorization algorithm. VIPE works in a multi-label mode where each text can belong to several categories.

The rest of the paper is described as follows: section 2 recalls related works on interactive multi-label classification, section 3 details the learning algorithm and experimental results on large datasets, section 4 presents the interactive process for a real life application. Our approach is currently tested by marketing service officers of the French telecom company Orange.

## 2  RELATED WORKS

The growing interest for personalized assistants in classification has led to the development of several interactive learning tools for various applications: e.g. cueFlick for image classification (Fogarty [10]), Smart Selection for file selection (Ritter [11]), CueTip for handwriting recognition (Shilman[14]), CueT for alarm classification (Amershi[13]), Regroup for user group creation in social networks (Amershi[12]). The first feedbacks reported in the papers are very encouraging but the vast majority of the current approaches are based on a single-label classification that forces items to span one label at a time. This simplified framework is not adapted to opinion mining where fine-grained analyses require multi-labelling as illustrated in the example above.

Multi-label classification has received significant attention over the past few years and a large number of algorithms have been proposed in the literature (Zhang [15]; Madjarov [18]; Tsoumakas[17]). Madjarov et al. [18] have conducted an extended comparison of twelve state-of-the-art algorithms and more recently Liu et al. [9] have relied upon this work to investigate multi-label classification approaches for sentiment analysis of microblogs. As far as we know, this study is the first in that domain. Its results are promising but their computation depends on predefined sentiment dictionaries which are not customized for the real-life data that users are interested in.

Combining multi-label classification and personalization via interactivity is a new challenge. The integration of a multi-label algorithm into a human-centered interactive system is hampered by a double constraint: learning from few training examples and in a limited time. Nair Benrekia et al. [19] have recently proposed an extensive comparative study of the behavior of multi-label learning algorithms in an interactive framework. Their experiments have highlighted the potentiality of a random forest based approach but they also have shown a degraded performance in terms of time computation for high-dimensional datasets. An alternative for reducing the computation time is to reduce the data dimensionality. Matrix factorization techniques are promising candidates for multi-label interactive classification. First versions for interactive classification, with penalty terms to take into account application-dependent constraints were reported in Berry [22] and recent studies have confirmed their interest in interactive content analysis (Bakharia [20]). In this paper, we have developed an adapted version of Funk's algorithm[†], also known as the Gravity algorithm (Takács [8]), which was initially applied to recommender systems. Our algorithm works on one-class sparse matrices where textual data are encoded by sets of n-grams. By discovering links between these data in a semi-supervised learning mode, it produces good quality results without important human effort for the pre-labelling.

## 3   MULTI-LABEL CLASSIFICATION WITH VIPE

Let us consider a set $\mathcal{U}$ of $m$ texts. Each text is automatically encoded by a set of n-grams. More precisely, we here retain all 1 to 3-grams decompositions (e.g for 3-grams "this is the", "is the best", … "best 4G network") that occur at least two times in $\mathcal{U}$. The set $\mathcal{G}$ of size $n_1$ contains all the n-grams associated with the $m$ texts.

At the beginning of the learning process, the user defines a set $\mathcal{L}$ of $n_2$ labels he/she interested in and he/she labels a small subset $\mathcal{T}$ of $m_1$ texts either positively or negatively. The remaining texts of $\mathcal{U}$ constitute the set $\mathcal{S}$ of $m_2$ unlabeled texts which is much larger than $\mathcal{T}$.

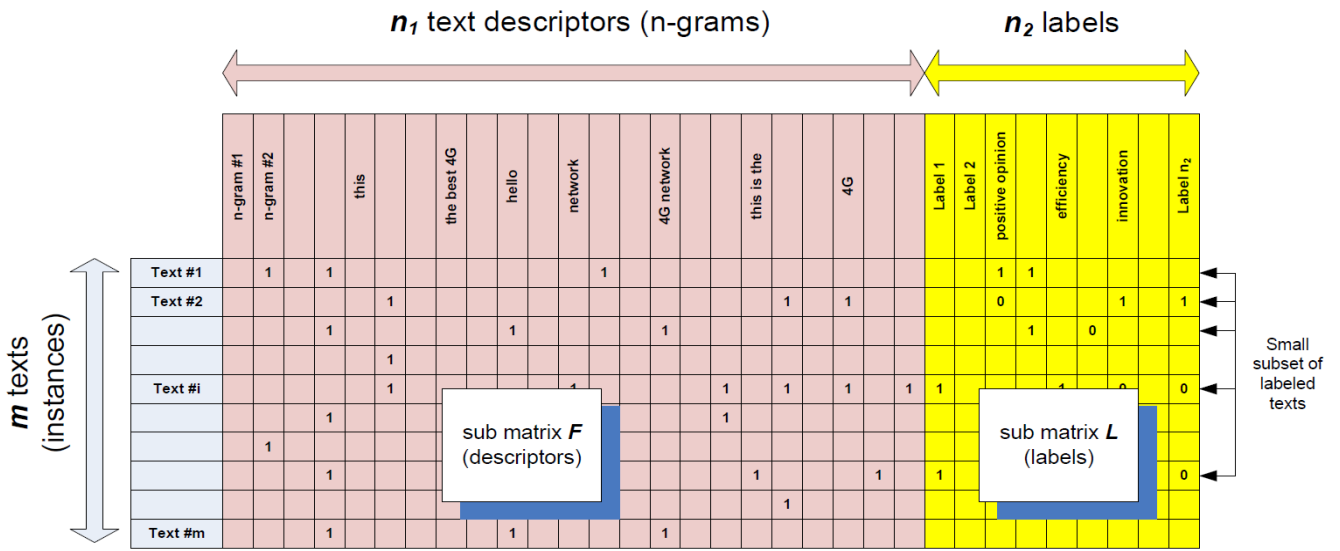

**Figure 1: The large sparse matrix X of the texts described by their n-grams and their labels.**

Let us denote by $X$ a $m \times n$ matrix which stores the set of $m = m_1 + m_2$ texts described by a set of $n = n_1 + n_2$ variables (n-grams and labels of the texts). The matrix $X$ can be decomposed into two sub-matrices $F \in \mathbb{R}^{m \times n_1}$ and $L \in \mathbb{R}^{m \times n_2}$. The matrix $F$ encodes the presence/absence of the n-grams for each text: $f_{ij} = 1$ if the $i^{th}$ text contains the $j^{th}$ n-gram of $\mathcal{G}$ and $f_{ij}$ is empty otherwise. This setting optimizes memory management. The matrix $L$ encodes the labels of the texts: $l_{ij} = 1$ (resp. 0) if the $i^{th}$ text is labeled positively (resp. negatively) with the $j^{th}$ label. If the $i^{th}$ text is not labeled with the $j^{th}$ label, the cell $l_{ij}$ remains empty. The matrix $L$ is also empty for the $m_2$ rows corresponding to the unlabeled texts of $S$. The goal of the matrix factorization in VIPE is to predict these empty rows i.e. the labels of the texts in $S$.

---

[†] http://sifter.org/~simon/journal/20061211.html

## 3.1 The fast factorization approach

The matrix factorization allows to approximate the sparse $m \times n$ matrix $X$ by a full low-rank matrix $\hat{X} = P^T Q$ where $P \in \mathbb{R}^{n \times k}$ and $Q \in \mathbb{R}^{m \times k}$ are respectively $k$ – dimensional ($k \ll n$ and $k \ll m$) latent representations of the rows and columns of $X$. The obtained low-rank matrix is used as an approximation of the missing values of $X$ (here the labels of the texts in $S$ that the user wants to predict).

The $P$ and $Q$ matrices are learned by minimizing the Root Mean squared Error (RMSE) which is the average square error $e_{ij}^2$ (error between the low-rank matrix approximation $\hat{x}_{ij}$ and the ground-truth value $x_{ij}$) for each non-empty cell of $X$. The RMSE yields an analytically easy to derive quadratic error function which can be efficiently optimized with a gradient descent algorithm:

$$RMSE_X = \sqrt{\frac{\sum_{x_{ij} \in \mathcal{N}_z(X)} (x_{ij} - \hat{x}_{ij})}{|\mathcal{N}_z(X)|}} \qquad (1)$$

with $\hat{x}_{ij} = \sum_{w=1}^{k} p_{iw} \times q_{wj}$, where $p_{iw}$ is the cell at the $i^{th}$ row and the $w^{th}$ column of $P$, and $q_{wj}$ is the cell at the $w^{th}$ row and the $j^{th}$ column of $Q$. $\mathcal{N}_z(X)$ is the set of non-empty cells in the matrix $X$.

The matrices $P$ and $Q$ are randomly initialized. Then, the factorization algorithm randomly considers all the non-zero cells of the matrix $X$: for each cell $x_{ij}$, it computes the gradient of the quadratic error $e_{ij}^2$ and it back-propagates it to update the factors $p_{i.}$ and $q_{.j}$. The gradient of $e_{ij}^2$ according is computed as follows:

$$\frac{\partial}{\partial p_{iw}} e_{ij}^2 = -2 \times e_{ij} \times q_{wj} \text{ and } \frac{\partial}{\partial q_{iw}} e_{ij}^2 = -2 \times e_{ij} \times p_{wj} \qquad (2)$$

To decrease the model prediction error and better approximate the values $x_{ij}$, the factors are updated in the opposite direction to the gradient:

$$p'_{iw} = p_{iw} + \alpha \times e_{ij} \times q_{wj} \text{ and } q'_{wj} = q_{wj} + \alpha \times e_{ij} \times p_{iw} \qquad (3)$$

where $\alpha$ is the learning rate (usually taking a small value i.e. $10^{-2}$). To prevent overfitting of the factor values, a regularization rate $\gamma$ (usually set around $8 \times 10^{-3}$) is added during the learning process:

$$p'_{iw} = p_{iw} + \alpha \times e_{ij} \times q_{wj} - \gamma \times p_{iw} \text{ and } q'_{wj} = q_{wj} + \alpha \times e_{ij} \times p_{iw} + \gamma \times q_{wj} \qquad (4)$$

The factor values are clipped into the interval $[-1; +1]$ to prevent divergence to infinity. Moreover, as the matrix factorization only considers the non-empty values here, $F$ can be confused with a matrix full of ones, which could lead to a nonsense approximation. To prevent that situation, the algorithm has to be informed that some of the empty cells carry a zero value. More precisely, during the learning phase, whenever a gradient is back propagated on a non-empty cell $f_{ij}$ of $F$, a gradient is also back propagated on an empty cell of $F$ randomly sampled.

Early stopping is used to avoid overfitting. At the end of each learning iteration, the RMSE is evaluated on a small validation set: if it does not decrease during a fixed number of iterations, the learning process stops and the latest matrices $P$ and $Q$ are considered as optimal.

At any step of the process, a real-valued prediction of the $j^{th}$ label for the $i^{th}$ text is stored in $\hat{X}$ at the $i^{th}$ row and the $(n_1 + j)^{th}$ column. In practice these values are used as scores to rank the most likely labels associated to an unlabeled text.

The complexity of the algorithm is linear with respect to the size of the matrix $X$ and is function of the number of passes (*NbPasses)* required for its convergence: it is equal to $O(n \times m \times NbPasses)$. However, experiments have shown that *NbPasses* is generally limited.

## 3.2 Validation of the algorithm on large matrices

To the best of our knowledge, there are not yet shared datasets for opinion multi-label classification and the ones used within the telecommunication company Orange cannot be detailed here for confidentiality constraints. For these reasons, we analyze the performance of the VIPE's fast factorization algorithm on classical datasets from the literature with characteristics close to data encountered in opinion mining (i.e. large one-class sparse matrices). The three chosen datasets (MovieLens, NetFlix and IMDB table movies × keywords, 2012 snapshot – see http://www.imdb.com/interfaces) are concerned with Video On Demand and their sizes are various: from 6000 to 330 000 for the number $n$ of lines (texts) and from 3600 to 480 000 for the number $m$ of columns (n-grams + labels). Let us note that the datasets MovieLens and NetFlix contain ratings. They have been transformed into $0/1$ values: ratings are recoded 1 if they are greater than the mean rating of the matrix and 0 otherwise.

Various measures have been introduced to evaluate the quality of the predictions (e.g. Madjarov [18]). For the experiments presented in this paper, we use the multi-label $BER$ (Balanced Error Rate) ratio which measures the number of incorrectly classified labels per instance:

$$BER = \frac{1}{n_S}\sum_{i=1}^{n_S} \frac{1}{2} \times \left(\frac{FP_i}{FP_i+TN_i} + \frac{FN_i}{FN_i+TP_i}\right) \quad (5)$$

where $TP_i$, $TN_i$, $FP_i$ and $FN_i$ are the number of respectively true positive, true negative, false positive and false negative labels in the $i^{th}$ row.

**Table 1: performance of the algorithm on large matrix**

| Dataset | one-class matrix | Number of columns $m$ | Number of rows $n$ | Number of logs (non-zero cells) | Average number of columns entries | Average number of rows entries | Zero value rates | BER |
|---|---|---|---|---|---|---|---|---|
| MovieLens 1M | no (ratings) | 3 699 | 6 040 | 1 000 202 | 148 | 243 | 96% | 11.3 ± 0.001% |
| NetFlix | no (ratings) | 480 039 | 17 770 | 90 430 940 | 188 | 5 088 | 99% | 12.9 ± 0.001% |
| IMDB | yes (movies × keywords) | 115 081 | 331 419 | 3 558 542 | 31 | 11 | 99.991% | 7.4± 0.005% |

All the tests were carried out with the same parameters: learning rate $\alpha = 0.001$; number of factors: $k = 16$ ; regularization parameter $\lambda = 0.008$. These parameters are issued from our large expertise with the Gravity algorithm. All the datasets were split into 10 folds (90% in train, 10% in test) with a classical 10-fold validation process.

The results for the BER measure which evaluates the misclassified labels are the following: 11.3% (Movie Lens), 12.9% (Netflix), 7.4% (IMDB). The experiments confirm that the learning process withstands the large data sparsity (more than 96% of missing values for the best cases). Details are given in Table 1.

## 4 APPLICATION

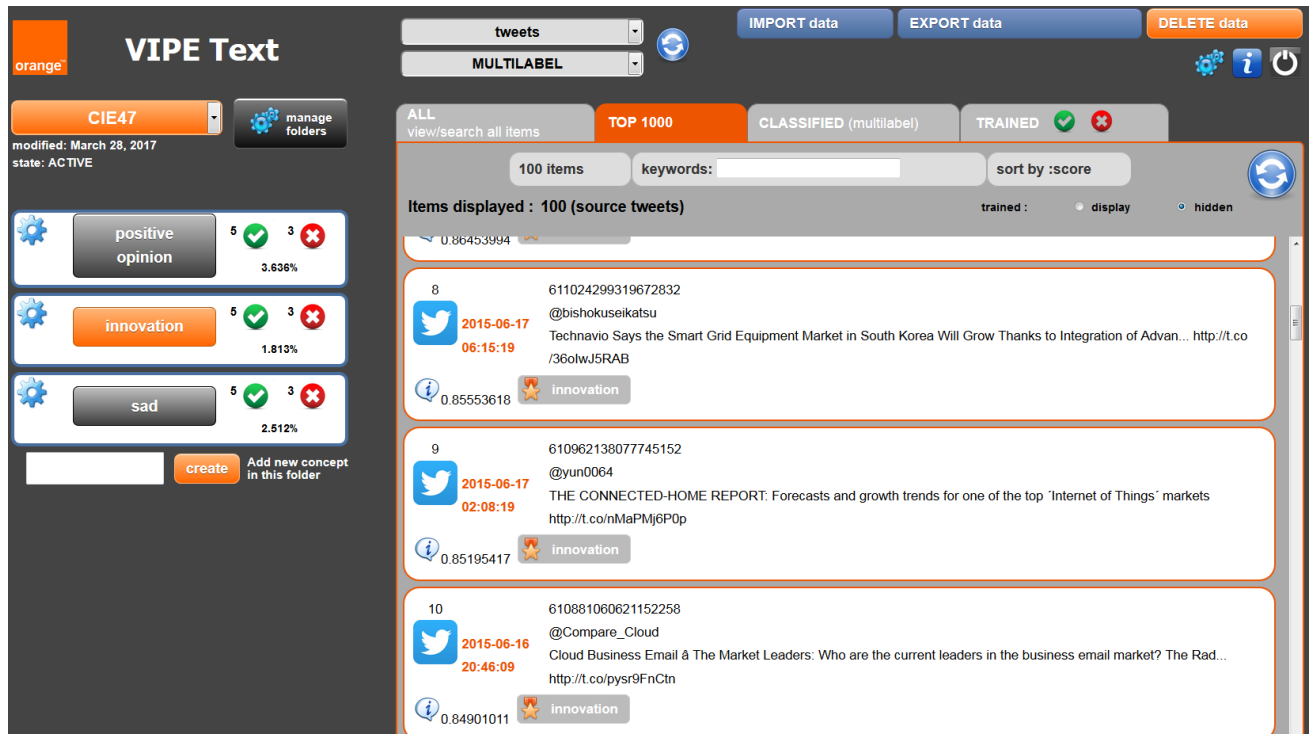

**Figure 2: VIPE interface: The left panel is used for label management (consult, add, delete). VIPE provides the texts with the best score for a selected label. The top panel is used for text import. The front panel allows text exploration: consult the texts, select them, annotate them, and visualize their predicted scores for the available labels. Additional tools are proposed for further analysis and export.**

For real-life applications VIPE has been integrated as a web-server application composed of 4 main modules:

1. a database manager which integrates the sparse matrix $X$ i.e. the texts representation $F$ and the label information $L$.
2. a module that uploads the text files and exports the classification after the training stage. All data sources are shared between users so that VIPE can learn via multiple sources but the learned labels are private. The size of the data stored in 2016 was: 446 700 texts, 1 380 000 n-grams, 156 labels.
3. a engine module which uses the matrix factorization algorithm described above to predict the scores for each label and each text.
4. a web-based user interface (figure 2) with several tools to easily manipulate the texts, annotate them, and manage the labels.

After importing data, users define their labels on the fly. With only a few interactions, VIPE is able to provide the label scores for any text of the corpora. If users are not satisfied with the results, they may propose a correction which is immediately taken into account by the learning algorithm.

After an interactive stage whose time depends on the data size, users feel confident with VIPE's proposals and can visualize the results (e.g top 1000 most suitable texts for a label and top labels for a text) and export them. In real-life situations where some online surveys contain hundreds of thousands of small texts of the customer feedbacks, less than two hours are necessary for the users to become confident in the algorithm predictions and to decide to export the results.

A user test has been carried out since the beginning of 2016 in several internal services of the telecommunication company Orange, mainly in marketing services, and about twenty people are currently using VIPE. From a qualitative point of view, users are very satisfied with its ease of use and its performances.

## 5 CONCLUSION

Classifying opinions expressed on a large scale in user-generated contents like social networks, Web surveys and forums is an important challenge for companies to adapt their relationships with their customers in a reactive way. This paper presents the new interactive multi-label classification approach VIPE to help users to classify large amounts of short texts. VIPE is already used by the telecommunication company Orange especially for survey analysis.

Stimulated by very promising feedbacks, improvements are planned in the short term. Our algorithm currently works with an unsupervised dimensionality reduction but we now are investigating a supervised approach to tackle the high dimensions in a learning process constrained by the response times. The objective is to be able to efficiently tackle matrices larger than $10^6 \times 10^6$ in our interactive framework. Recently, a comparative study (Pacharawongsaka [16]) has highlighted the interest of a dual reduction which simultaneously reduces the descriptor space and the label space. Integrating such a dual reduction in VIPE is a promising direction for enhancing the VIPE's engine.

## 6 ACKNOWLEDGMENTS

The authors would like to thank Sylvie Tricot (research engineer at Orange Labs Lannion) for her contribution to the implementation of VIPE.

## 7 REFERENCES

[1] **Gundecja, P., Liu, H. 2012.** Mining social media: a brief introduction, *INFORMS*, vol. 9, n°4, pp. 1-17.

[2] **Power, D., Philips-Wrem, G. 2011.** Impact of social media and Web 2.0 on decision making, Journal of Decision Systems, vol. 20, n°3, pp. 249-261.

[3] **Piryani, R., Madhavi, D., Singh, V.K. 2017.** Analytical mapping of opinion mining and sentiment analysis research during 2000-2015, Information Processing & Management, vol. 53, n°1, pp. 122-150.

[4] **Ravi, K., Ravi, V. 2015.** A survey on opinion mining and sentiment analysis: tasks, approaches and applications, Knowledge-based systems, vol. 89, pp. 14-46.

[5] **Keim, D. A., Krstajic, M., Rohrdantz, C., Schreck,T. 2013.** Real-time visual analytics for text streams. Computer (7), 47–55.
[6] **Jordan, M.I., Mitchell, T.M. 2015.** Machine learning: trends, perspectives and prospects, Science, vol. 649 (6254), pp. 255-260.
[7] **Takács, G., Pilászy, I., Németh, B., Tikk, D. 2008.** Investigation of various matrix factorization methods for large recommender systems. ICDM Workshops, pp. 553-562.

[8] **Sun, S., Luo, C., Chen, J. 2017.** A review of natural langage processing techniques for opinion mining systems, Information Fusion, vol. 36, pp. 10-25.

[9] **Liu, S.M., Chen, J.-H. 2015.** A multi-label classification based approach for sentiment classification. Expert Systems with Applications, vol. 42, pp. 1083-1093.

[10] **Fogarty, J., Tan, D., Kapoor, A., and Winder, S. (2008).** Cueflik: interactive concept learning in image search. In Proceedings of the SIGCHI Conference on Human Factors in Computing Systems, pages 29–38. ACM.

[11] **Ritter, A. and Basu, S. (2009).** Learning to generalize for complex selection tasks. In Proceedings of the 14th international conference on Intelligent user interfaces, pages 167–176. ACM.

[12] **Amershi, S., Fogarty, J., Weld, D. (2012).** Regroup: Interactive machine learning for ondemand group creation in social networks. In Proceedings of the SIGCHI Conference on Human Factors in Computing Systems, pages 21–30. ACM.

[13] **Amershi, S., Lee, B., Kapoor, A., Mahajan, R., Christian, B. (2011).** Cuet: human guided fast and accurate network alarm triage. In Proceedings of the SIGCHI Conference on Human Factors in Computing Systems, pages 157–166. ACM.

[14] **Shilman, M., Tan, D. S., Simard, P. (2006).** Cuetip: a mixed-initiative interface for correcting handwriting errors. In Proceedings of the 19th annual ACM symposium on User interface software and technology, pages 323–332. ACM.

[15] **Zhang, M. L., Zhou, Z. H. (2014).** A review on multi-label learning algorithms. IEEE transactions on knowledge and data engineering, 26(8), 1819-1837.

[16] **Pacharawongsakda, E., Theeramunkong, T. (2016).** A Comparative Study on Single and Dual Space Reduction in Multi-label Classification. In Knowledge, Information and Creativity Support Systems: Recent Trends, Advances and Solutions (pp. 389-400). Springer International Publishing.

[17] **Tsoumakas, G., Katakis, I. (2007).** Multi-label classification: An overview. International Journal of Data Warehousing and Mining (IJDWM), 3(3) :1–13.

[18] **Madjarov, G., Kocev, D., Gjorgjevikj, D., Dzeroski, S. (2012).** An extensive experimental comparison of methods for multi-label learning. Pattern recognition, 45(9) :3084-3104.

[19] **Nair-Benrekia N. Y., Kuntz P., Meyer F. (2015).** Learning from multi-label data with interactivity constraints: an extensive experimental study. Expert Systems with Applications, vol. 42, n° 13, pp. 5223-5736.

[20] **Bakharia, A., Bruza, P., Watters, J., Narayan, B., Sitbon, L. (2016, March).** Interactive Topic Modeling for aiding Qualitative Content Analysis. In Proceedings of the 2016 ACM on Conference on Human Information Interaction and Retrieval (pp. 213-222). ACM.

[21] **Berry, M. W., Browne, M., Langville, A. N., Pauca, V. P., Plemmons, R. J. (2007).** Algorithms and applications for approximate nonnegative matrix factorization. Computational statistics & data analysis, 52(1), 155-173.